\documentclass[twocolumn,showpacs,preprintnumbers,amsmath,amssymb,prb,floatfix]{revtex4}

\usepackage{graphicx}
\usepackage{dcolumn}
\usepackage{bm}

\begin{document}

\title{Two Dimensional Spin-Polarized Electron Gas at the Oxide Interfaces
}

\author{B. R. K. Nanda}
\author{S. Satpathy}%
\affiliation{%
Department of Physics $\&$ Astronomy, University of Missouri, Columbia, MO 65211}%

\date{\today}

\begin{abstract}

The formation of a novel spin-polarized 2D electron gas at the LaMnO$_3$ monolayer embedded in SrMnO$_3$ is predicted from the 
first-principles density-functional calculations. The  La (d) electrons become confined in the direction normal to the interface in the potential well of the La layer, serving as
a positively-charged layer of electron donors.
These  electrons mediate a
ferromagnetic alignment of the Mn t$_{2g}$ spins near the interface via the Anderson-Hasegawa double exchange and become, in turn,
spin-polarized due to the internal magnetic fields of the Mn moments. 
\end{abstract}

\maketitle


Recent advances in the fabrication of high-quality epitaxial interfaces between perovskite oxides have led to a rapid surge of interest in the study of new interface electronic states.
A number of oxide interfaces have been shown to possess electrons confined to
the interface region forming a two-dimensional electron gas (2DEG). A clear example is an isolated [100] monolayer of
LaTiO$_3$ grown in a SrTiO$_3$ host \cite{ohtomo,popovic,hamann,millis}. The trivalent La substituting for the divalent Sr in essence behaves as
a positively-charged layer of electron-donor dopants producing a wedge-shaped potential at the interface, where the electrons become confined. A similar type of electron gas has been
observed at the much studied LaAlO$_3$/SrTiO$_3$ interface \cite{hujiben}, although the exact origin of the electron gas there remains controversial. A somewhat different physics resulting from the
interface polarization charges produces the 2DEG
at the nitride and the oxide interfaces such as GaN/Al$_x$Ga$_{1-x}$N \cite{hang} and ZnO/ Mg$_x$Zn$_{1-x}$O \cite{ohno}. These electron gases often show clear Shubnikov-de Haas oscillations and even superconductivity has been observed in one of the systems just recently.\cite{Reyren}  

\begin{figure}
\includegraphics[width=6.5cm]{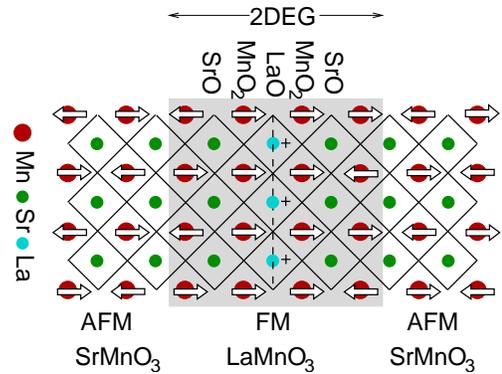}
\caption{\label{sketch} (color online) The manganite interface structure with a monolayer of LaMnO$_3$ embedded in the SrMnO$_3$ bulk, with the shadowed region indicating the spin-polarized 2DEG confined to the
interface region. Oxygen atoms occur at the intersections of the checkered lines and form MnO$_6$ octahedra around the Mn atoms.}
\end{figure}

Since many of these oxides contain magnetic atoms as well, the question arises as to whether one may be able to get a spin-polarized electron gas at the interface, using the internal magnetic fields due to these magnetic atoms. In this Letter, we predict from density-functional band calculations the existence of just such a phase at the manganite interface structure consisting of  a LaMnO$_3$ (LMO) monolayer  embedded in the SrMnO$_3$ (SMO) bulk, sketched in Fig. \ref{sketch}. 


The results presented here are obtained from density-functional
 (DFT) studies of the (LMO)$_{1}$/(SMO)$_{7}$ layered superlattice using the
linear augmented plane wave (LAPW) \cite{wien} 
and the linear muffin-tin orbitals method (LMTO)\cite{lmto} with generalized gradient approximation (GGA)\cite{gga} or the 
Coulomb-corrected local spin density approximation (LSDA+U) for the exchange-correlation potential.
The supercell consisted of twice this formula unit because of the 
magnetic structures considered in the paper.
The structural relaxation was performed using the LAPW-GGA
method for the magnetic structure. The calculated cubic lattice constant for the La compound
is 3.915 \AA\ and for the Sr compound is 3.802 \AA, which are close
to the experimental values of 3.935 \AA\ and 3.805 \AA\ respectively.
For the purpose of relaxation, we fixed the in-plane lattice parameter 
of the superlattice to be 3.802 \AA\ corresponding to the
bulk lattice constant of the Sr compound. This was also the 
lattice constant for the SMO part in the direction normal to the superlattice,
while the same for the LMO monolayer was taken to be  4.15 \AA, which conserves the volume of the LMO unit cell in the bulk. The atoms were then relaxed along the c-axis as the symmetry prohibits their motion along the plane. 

\begin{figure}
\includegraphics[width=8cm]{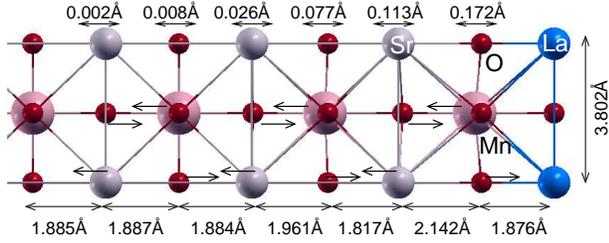}
\caption{ (color online) Relaxed atomic positions for the (LMO)$_1$/(SMO)$_7$ superlattice. The atomic displacements, shown by arrows,
indicate that the cation-anion polarizations diminish quickly as one goes away from the interface.}
\label{relaxed} 
\end{figure}

      The results, presented in Fig. \ref{relaxed}, indicates the movement of the cations away from the interface, while the anions move towards the interface due to the electrostatic attraction with the positively-charged La layer. This is similar to the cation-anion polarization  obtained for the (SrTiO$_3)_n$/(LaTiO$_3)_1$ heterostructure, where there also exists at the interface a
      positively-charged layer of La atoms.\cite{hamann,millis,larson}
The electronic structure for the relaxed lattice was obtained using the LMTO method with the LSDA+U approximation, using the Coulomb and the exchange interaction parameters of U = 5 eV and J = 1 eV.

\begin{figure}
\includegraphics[width=5cm]{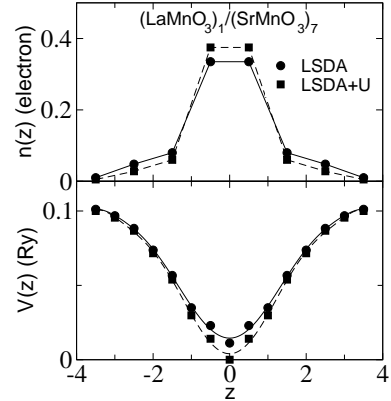}
 \caption{The cell-averaged potential $V(z)$ calculated from Eq. (\ref{Eq-potential}) with distance
$z$ from the La layer in units of SrMnO$_3$ monolayer thickness (bottom), and 
the donated electrons from La, one per atom, occupying the e$_g$ 
orbitals of the adjacent Mn atoms (top).  The layer occupancy $n(z)$
is the occupancy of electrons for the individual MnO$_2$ layers
and they have predominantly 
Mn e$_g$ character. }
\label{potential} 
\end{figure}

The magnetic ground states of the two bulk compounds are very different. While SMO with its Mn$^{4+}$ (t$_{2g}^3$e$_{g}^0$)
configuration is a G-type antiferromagnet (AFM) with the magnetic interaction between the Mn core spins driven by superexchange\cite{millis1}, LMO is an A-type antiferromagnet with Mn$^{3+}$ (t$_{2g}^3$e$_{g}^1$) electronic
configuration, with the partially-filled e$_g$ electrons mediating a
ferromagnetic double exchange between the t$_{2g}$ core spins on the
MnO$_2$ planes\cite{popovic1,feinberg}. For the present interface, there is just one extra 
electron per La atom, which is localized near the interface, occupying
the Mn e$_g$ orbitals (see Fig. \ref{potential}). These electrons serving as the itinerant
electrons in the standard double exchange picture\cite{hase} are expected to modify the magnetism of
the Mn t$_{2g}$ core spins at the interface.

In order to study the magnetic ground state at the interface, we have
considered several magnetic configurations and computed their total
energies. We find the lowest-energy structure to be the one shown in Fig. \ref{sketch}, where the two MnO$_2$ layers on either side of the La layer are ferromagnetic, while the remaining Mn atoms retain the
N{\'e}el G-type AFM of the SMO bulk.  The rest three structures 
that we examined all have higher energies, viz., (i) the structure with
a complete G-type magnetism, (ii) one where the layer ferromagnetism is extended up to the second MnO$_2$ layer on either side of the interface, and (iii) the structure with the two MnO$_2$ layers across the interface aligned antiferromagnetically, rather than ferromagnetically as in Fig. \ref{sketch}. The layer ferromagnetic alignment at the interface is explained by  the fact that the itinerant
e$_g$ electrons mediating the double exchange reside predominantly in the two layers adjacent
to the interface, even though a small e$_g$ charge spreads to 
layers beyond the first layer as seen from Fig. \ref{potential}. 

We note that the magnetic structure
of the interface is very much dependent on the strain condition.\cite{tokura,strainbrk}
 In the present work, the 
lattice constant along the interface corresponds to the SMO bulk lattice. For a strong enough tensile or compressive strain, the
layer ferromagnetic structure shown in Fig. \ref{sketch}  is no longer energetically favorable;\cite{strainbrk} Consequently, the spin-polarized
2DEG will not exist for these strain conditions.

The variation of the potential seen by the electron at the interface may be calculated from the variation of some reference energy in the LMTO calculation. A convenient reference energy is the cell-averaged point-charge Coulomb potential V, shown in Fig. \ref{potential}, which was calculated by first averaging the potential over the volume of the {\it i}-th Wigner-Seitz atomic sphere:\cite{lamb,popovic}
\begin{equation}
V_i = \frac{3q_i}{2s_i} + \sum_j\frac{q_j}{|r_i-r_j|}
\label{Eq-potential}
\end{equation}
and then by averaging over all spheres with a weight factor proportional to their volumes:     $V= \sum_i\Omega_iV_i/\sum_i\Omega_i$, where $\Omega_i = 4\pi s_i^3 /3 $ is the sphere volume, $s_i$ is the sphere radius, $r_i$ its position, and $q_i$ is the total charge, nuclear plus electronic. In Eq. (1), the first term is the sphere average of the potential of the point charge located at the center of the muffin-tin sphere and the second term is the Madelung potential due to all other spheres in the solid. 

The results plotted in Fig. \ref{potential} shows the screening of the
bare linear potential due to the electrostatic field of the charged La plane caused by the electronic as well as the lattice polarization. The screened potential is deep enough to localize the donor electron within just a few layers of the interface. In fact, we find that about 0.7 e$^-$ is located on the first MnO$_2$ layers, 0.14 e$^-$ on the second layers, and the remaining 0.16 e$^-$ is spread between the remaining atoms.


The presence of a substantial amount of the e$_g$ charge on the first MnO$_2$ layer is consistent with the double exchange mechanism for the layer ferromagnetism (Fig. \ref{sketch}) found from the DFT calculations. However, although in the DFT results, the structure, where the second MnO$_2$ layer is also ferromagnetic in addition to the first, was not energetically favorable, the leaked e$_g$ electrons into the second layer could result in a canted ferromagnetic state for this layer. Whether a canted state forms or the antiferromagnetism is retained depends on the amount of charge leakage into this layer and the strength of the double exchange interaction.

\begin{figure}
\includegraphics[width=5.3cm]{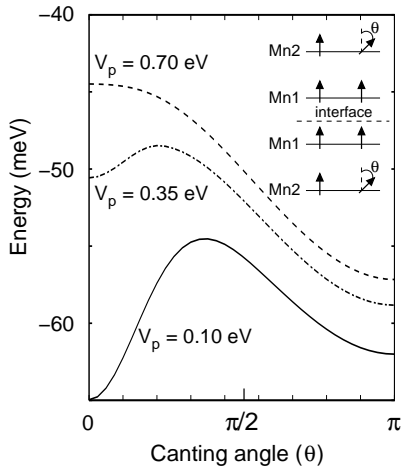}
\caption{Total energy obtained from Eq. \ref{model} as a function of the canting angle between the nearest neighbor Mn spins in the second MnO$_2$ layer  for different values of the potential  V$_p$.  
A magnitude of $V_p \approx 0.5$ eV, which results in roughly the same
electron leakage  into the second layer as obtained from the DFT, yields an antiferromagnetic ground state as inferred from the figure, indicating the absence of a
canted state.}
\label{canting}
\end{figure}

         To address this issue, we have studied the Anderson-Hasegawa double-exchange model\cite{hase} on a four-layer lattice
         (Fig. \ref{canting}, inset), each layer being a square lattice, as appropriate for the MnO$_2$ layers.
         The model Hamiltonian, restricted to the Mn sites up to second MnO$_2$ planes away from the interface,   is given by

\begin{eqnarray}
\nonumber H =  \sum_{i\sigma} \epsilon_{i\sigma} n_{i\sigma} + t\sum_{\langle ij \rangle \sigma} c^{\dagger}_{i\sigma}c_{j\sigma} + H.c\\
+ J\sum_{\langle ij \rangle} \hat{S_i}.\hat{S_j} - 2J_H\sum_i  \vec{S_i},\vec{s_i},
\label{model}
\end{eqnarray}   
and it describes the double exchange interaction of the itinerant electrons in a lattice of localized spins.
 Here, $c^{\dagger}_{i\sigma}$, $c_{i\sigma}$ are the field operators for the e$_g$ carriers, treated within a one-band model, with $i, \sigma$ being the site and spin indices, $\vec S_i$ is the localized t$_{2g}$ spin  and $\vec s_i$ = $\frac{1}{2}\sum_{\mu\nu}c^{\dagger}
_{i\mu}\vec{\tau}_{\mu\nu}c_{i\nu}$ is the itinerant spin electron density, with $\vec{\tau}$ being the Pauli spin matrices.
The onsite energy $\epsilon_{i \sigma}$  ($V_p$ on the second layer and zero on the first) 
describes the electric field at the interface and is the parameter that controls the leakage of the itinerant electron into the second layer.
 J is  the superexchange interaction between the localized spins,
 while $J_H$ is the Hund's coupling between the itinerant and the localized electrons. Guided by the earlier DFT calculations \cite{popovic1, pickett, hakim} the typical values of the  parameters  are:
$t \sim -0.15$ eV, $J \sim 7 $ meV, and $J_H \sim 1 $eV. Note that unlike our earlier work,\cite{Satpathy-canting} here we neglect the on-site Coulomb energy between the itinerant carriers, since the number of carriers is small.

The Hamiltonian (\ref{model}) is solved by diagonalizing 
a $16 \times 16$  Hamiltonian matrix (eight Mn atoms per unit cell
and two spin types) for a number of $\vec k$ points in the two-dimensional Brillouin zone and the the total energy is calculated by summing over the occupied states
(two electrons per cell). The results summarized in Fig. \ref{canting}
show that an antiferromagnetic second layer  (canting angle
$\theta = \pi$) is overwhelmingly favored over a canted state for a parameter of  $V_p \approx 0.5$ eV, which yields roughly the same amount of 
electron leakage into the second MnO$_2$ layer as obtained from the DFT calculations.


\begin{figure}
\includegraphics[width=7.3cm]{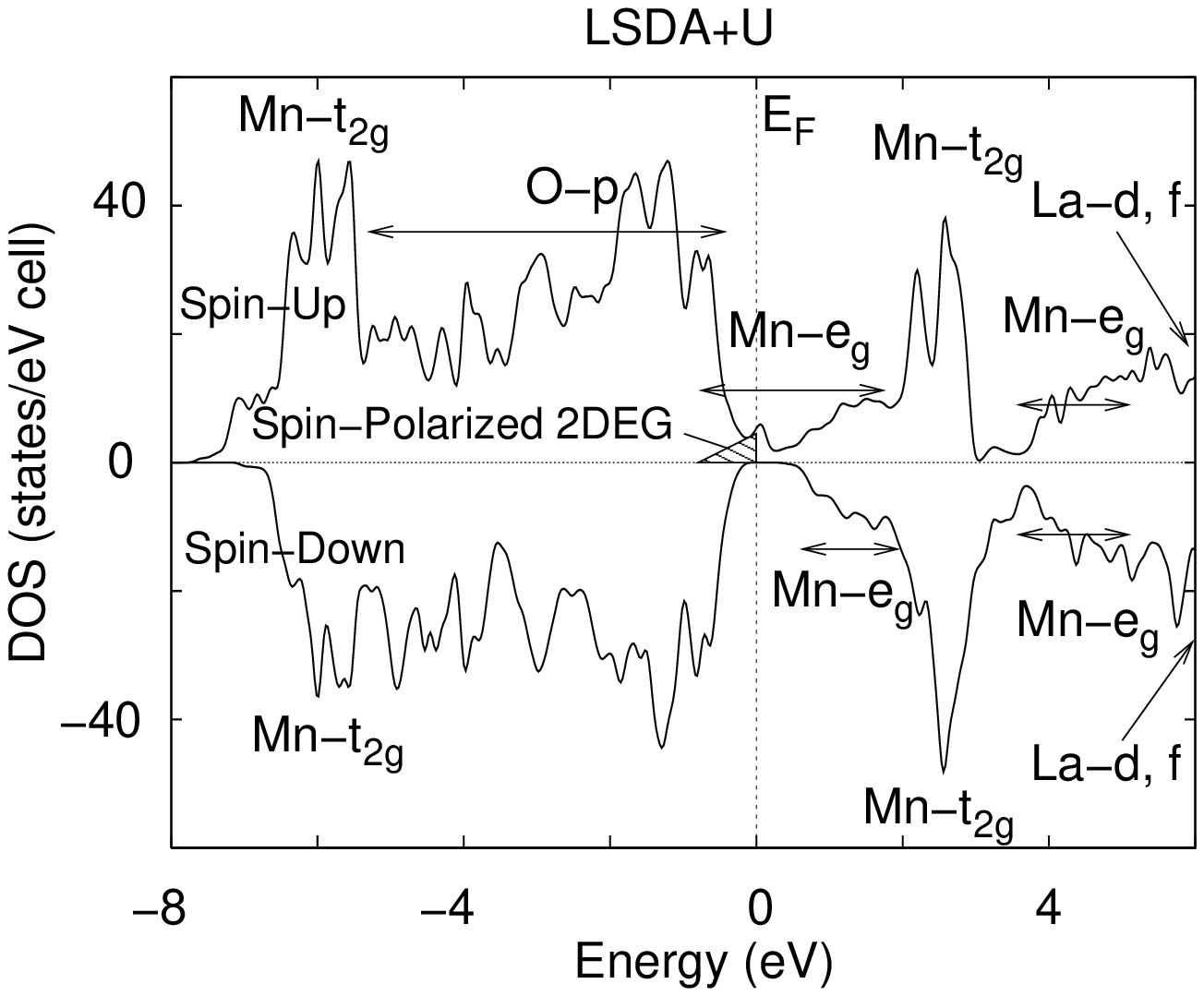} 
\includegraphics[width=8.3cm]{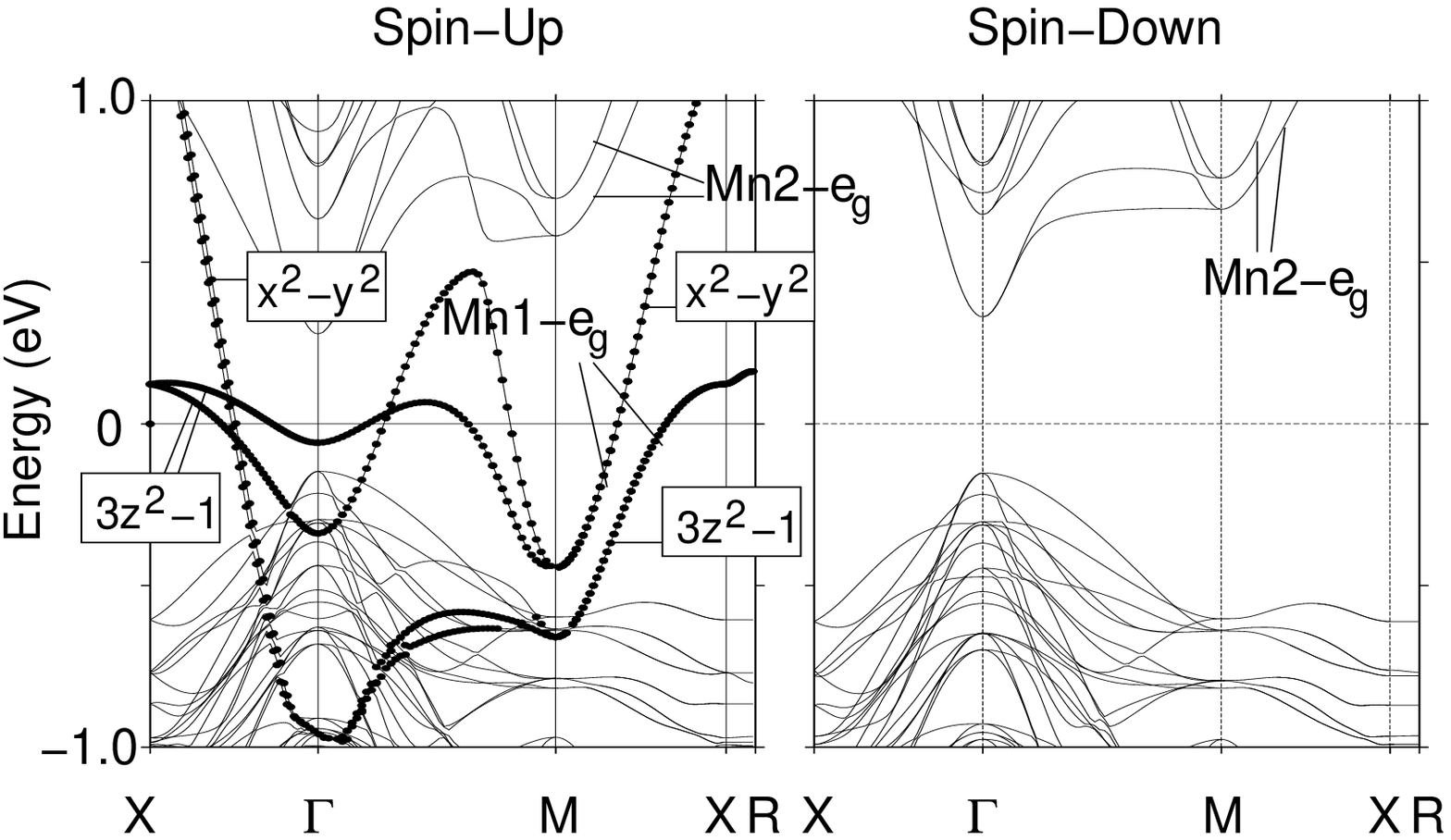}
\caption{ Total densities of states for the (LMO)$_1$/(SMO)$_7$ interface for the majority and the minority spins (top) and the
band structure in the vicinity of the Fermi energy (bottom). The symmetry points are: $\Gamma$ (0, 0, 0), X $(1, -1, 0)$, M $(0, 2, 0)$ and R $(1, -1, -2a/c)$ in units of $\pi /2a$, $a$ being the lattice constant along the plane of the interface. The symbols Mn1 and Mn2 indicate the first and the second layer atoms next to the interface.}
\label{bands}
\end{figure}

The electronic structure corresponding to the lowest-energy magnetic structure (Fig. \ref{sketch}) is shown in Fig. \ref{bands}.
The t$_{2g}$ states of one spin are occupied for each Mn atom and lie far below the Fermi energy $E_F$ because of
the octahedral crystal field produced by the MnO$_6$ octahedron and the strong Coulomb repulsion $U$. The vicinity of $E_F$ is occupied by the Mn-e$_g$ states. In the spin-majority channel, we see extended e$_g$ states crossing the Fermi level, while in the spin minority channel, they are unoccupied and open up a gap at the Fermi energy, so that we have a half metallic system.

While there are a number of half metallic systems known, the classic one being Fe$_3$O$_4$\cite{Zhang-fe3o4}, the present case is quite unique
in the sense that the electrons at the Fermi energy are confined
to the interface region, producing a spin-polarized 2DEG. This is in fact a key point of the paper.

 The bands crossing the Fermi energy are the majority-spin Mn1-e$_g$   (x$^2$-y$^2$ and 3z$^2$-1) states belonging to the first-layer manganese  atoms with bonding interaction across the interface. The corresponding anti-bonding states as well as all  minority-spin Mn1-e$_g$ states occur higher in energy and outside the energy range of Fig. \ref{bands}. In contrast to this, since the second layer manganese moments are antiferromagnetically organized, the Mn2-e$_g$  states
 occur both in the minority and majority spin channels as seen from the
 figure.
  
 As discussed earlier, the partially occupied e$_g$ states  mediate a ferromagnetic double exchange interaction between the t$_{2g}$ 
 core spins, which competes with the antiferromagnetic superexchange.
        This double exchange interaction is directional in nature in the sense that its strength in the xy-plane or along the z-axis depends on the occupancy of the individual e$_g$ orbitals (x$^2$-y$^2$ and 3z$^2$-1)\cite{strainbrk, tokura}. Since we have both orbitals significantly occupied, this leads to a strong double exchange both in the first MnO$_2$ layers and between these layers across the interface, resulting in the layer ferromagnetic structure as shown in  Fig. \ref{sketch}. 

\begin{figure}
\includegraphics[width=4.3cm]{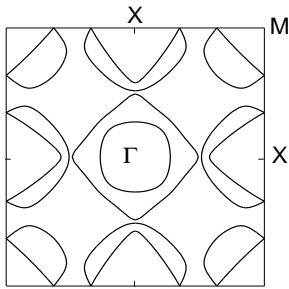}
\caption{The Fermi surface of (LMO)$_1$/(SMO)$_7$ shown in the interface Brillouin zone. }
\label{fermi} 
\end{figure}

  The interface Fermi surface, shown in Fig. \ref{fermi}, is constituted out of the Mn1-e$_g$   (x$^2$-y$^2$ and 3z$^2$-1) states and their orbital characters may be inferred from the band structure shown in Fig. \ref{bands}.  The Fermi surface consists of electron-like pockets at the
  $\Gamma$ and M points and hole-like pockets centered at the X points.  The holes have relatively higher mass, so that the transport along the interface may be expected to be electron like.
   
   Although the interface suggested in this paper has not been grown to our knowledge (but is certainly possible to grow), it is encouraging that there are several experimental works on the LMO/SMO superlattices, which seem to support the existence of 
   a ferromagnetic state at the interface.\cite{may,smadici,tokura}  It
   would be gratifying if the spin-polarized 2DEG can be established experimentally in these oxide systems.
   
In summary, from density-functional studies, we have predicted the formation of a spin-polarized 2DEG at the LaMnO$_3$
layer embedded in a thick SrMnO$_3$ bulk. This occurs due to the
confinement of the La electrons near the interface because of the electrostatic
potential of the positively-charged La layer. These electrons occupy the Mn-e$_g$ states near the interface,
mediating a double exchange interaction to stabilize a layer ferromagnetic structure of the Mn spins and become, in turn, completely spin-polarized due to the magnetic fields of the Mn atoms.
The N{\'e}el G-type antiferromagnetism of the bulk  SrMnO$_3$ is retained
in the second MnO$_2$ layer from the interface and beyond.
The complete spin-polarization of the electron gas   without any external magnetic field  is a new feature for the perovskite oxide interfaces. 

We  acknowledge support of this work by the U. S. Department of Energy through Grant No. DE-FG02-00ER45818.


\end{document}